\documentclass[twocolumn,english,aps]{revtex4}
\usepackage[T1]{fontenc}
\usepackage[latin9]{inputenc}
\usepackage{mathrsfs}
\usepackage{graphicx}
\usepackage{esint}
\usepackage{verbatim}   
\usepackage{color}      
\usepackage{subfigure}  
\usepackage{hyperref}   
\usepackage{array}
\usepackage{multirow}
\usepackage{babel}

\makeatletter

\let\jnl@style=\rm
\def\ref@jnl#1{{\jnl@style#1}}
\def\aj{\ref@jnl{AJ}}                   

\providecommand{\tabularnewline}{\\}

\@ifundefined{textcolor}{}
{%
 \definecolor{BLACK}{gray}{0}
 \definecolor{WHITE}{gray}{1}
 \definecolor{RED}{rgb}{1,0,0}
 \definecolor{GREEN}{rgb}{0,1,0}
 \definecolor{BLUE}{rgb}{0,0,1}
 \definecolor{CYAN}{cmyk}{1,0,0,0}
 \definecolor{MAGENTA}{cmyk}{0,1,0,0}
 \definecolor{YELLOW}{cmyk}{0,0,1,0}
}

\let\jnl@style=\rm
\def\ref@jnl#1{{\jnl@style#1}}

\def\aj{\ref@jnl{AJ}}                   
\def\actaa{\ref@jnl{Acta Astron.}}      
\def\araa{\ref@jnl{ARA\&A}}             
\def\apj{\ref@jnl{ApJ}}                 
\def\apjl{\ref@jnl{ApJ}}                
\def\apjs{\ref@jnl{ApJS}}               
\def\ao{\ref@jnl{Appl.~Opt.}}           
\def\apss{\ref@jnl{Ap\&SS}}             
\def\aap{\ref@jnl{A\&A}}                
\def\aapr{\ref@jnl{A\&A~Rev.}}          
\def\aaps{\ref@jnl{A\&AS}}              
\def\azh{\ref@jnl{AZh}}                 
\def\baas{\ref@jnl{BAAS}}               
\def\bac{\ref@jnl{Bull. astr. Inst. Czechosl.}}
\def\caa{\ref@jnl{Chinese Astron. Astrophys.}}
\def\cjaa{\ref@jnl{Chinese J. Astron. Astrophys.}}
\def\icarus{\ref@jnl{Icarus}}           
\def\jcap{\ref@jnl{J. Cosmology Astropart. Phys.}}
\def\jrasc{\ref@jnl{JRASC}}             
\def\memras{\ref@jnl{MmRAS}}            
\def\mnras{\ref@jnl{MNRAS}}             
\def\na{\ref@jnl{New A}}                
\def\nar{\ref@jnl{New A Rev.}}          
\def\pra{\ref@jnl{Phys.~Rev.~A}}        
\def\prb{\ref@jnl{Phys.~Rev.~B}}        
\def\prc{\ref@jnl{Phys.~Rev.~C}}        
\def\prd{\ref@jnl{Phys.~Rev.~D}}        
\def\pre{\ref@jnl{Phys.~Rev.~E}}        
\def\prl{\ref@jnl{Phys.~Rev.~Lett.}}    
\def\pasa{\ref@jnl{PASA}}               
\def\pasp{\ref@jnl{PASP}}               
\def\pasj{\ref@jnl{PASJ}}               
\def\rmxaa{\ref@jnl{Rev. Mexicana Astron. Astrofis.}}%
\def\qjras{\ref@jnl{QJRAS}}             
\def\skytel{\ref@jnl{S\&T}}             
\def\solphys{\ref@jnl{Sol.~Phys.}}      
\def\sovast{\ref@jnl{Soviet~Ast.}}      
\def\ssr{\ref@jnl{Space~Sci.~Rev.}}     
\def\zap{\ref@jnl{ZAp}}                 
\def\nat{\ref@jnl{Nature}}              
\def\iaucirc{\ref@jnl{IAU~Circ.}}       
\def\aplett{\ref@jnl{Astrophys.~Lett.}} 
\def\apspr{\ref@jnl{Astrophys.~Space~Phys.~Res.}}
\def\bain{\ref@jnl{Bull.~Astron.~Inst.~Netherlands}} 
\def\fcp{\ref@jnl{Fund.~Cosmic~Phys.}}  
\def\gca{\ref@jnl{Geochim.~Cosmochim.~Acta}}   
\def\grl{\ref@jnl{Geophys.~Res.~Lett.}} 
\def\jcp{\ref@jnl{J.~Chem.~Phys.}}      
\def\jgr{\ref@jnl{J.~Geophys.~Res.}}    
\def\jqsrt{\ref@jnl{J.~Quant.~Spec.~Radiat.~Transf.}}
\def\memsai{\ref@jnl{Mem.~Soc.~Astron.~Italiana}}
\def\nphysa{\ref@jnl{Nucl.~Phys.~A}}   
\def\physrep{\ref@jnl{Phys.~Rep.}}   
\def\physscr{\ref@jnl{Phys.~Scr}}   
\def\planss{\ref@jnl{Planet.~Space~Sci.}}   
\def\procspie{\ref@jnl{Proc.~SPIE}}   

\makeatother

\begin{document}
\title{Gravitational-wave sources from mergers of binary black-holes\\
catalyzed by fly-bys interactions in the field}
\author{Erez Michaely$^{*,1}$ \& Hagai B. Perets$^{2}$}
\affiliation{$^1$ Astronomy Department, University of Maryland, College Park, MD 20742\\
$^2$ Physics Department, Technion - Israel Institute of Technology, Haifa 3200004, Israel\\}
\email{erezmichaely@gmail.com}

\begin{abstract}
Several scenarios were suggested for the origins of gravitational-wave
(GW) sources from mergers of stellar binary black holes (BBHs). Here
we propose a novel origin through catalyzed formation of GW-sources
from ultra-wide binaries in the field. Such binaries experience perturbations
from random stellar fly-bys which excite their eccentricities. Once
a wide-binary is driven to a sufficiently small peri-center approach,
GW-emission becomes significant, and the binary inspirals and merges.
We derive an analytic model and verify it with numerical calculation
to compute the merger rate to be $\sim10{\rm \times f_{wide}\,{\rm Gpc}^{-3}yr^{-1}}$
($f_{{\rm wide}}$ is the fraction of wide BH-binaries), which is
comparable to the observationally inferred rate. The observational
signatures from this channel include spin-orbit misalignment; preference
for high mass-ratio BBH; preference for high velocity-dispersion host-galaxies;
and a uniform delay-time distribution. 
\end{abstract}
\maketitle
\section{Introduction}

Extensive theoretical studies over the past few decades have proposed
the existence of gravitational-wave (GW) sources arising from the
mergers of two compact objects, and provided a wide range of predicted
production rates of such sources \citep[e.g.][and more]{Belczynski2002,Belczynski2004,Belczynski2007,Belczynski2008,Belczynski2016,deMink2015,Dominik2012,Antonini2012,Antonini2014,Antognini2014,Petrovich2017}.
Observationally, eleven confirmed GW mergers have been detected by
aLIGO and VIRGO since their initial operation. These include $9$
mergers of binary black-holes (BBHs) and a single merger from a binary
neutron-star (NS)\citet{TheLIGOScientificCollaboration2018}. The
currently inferred BBH-merger rate from these observations (in the
local Universe) is $\mathscr{R_{{\rm BBH}}}=9.7-101{\rm Gpc^{-3}yr^{-1}}$;
while the merger rates of binary neutron-star is $\mathscr{R_{{\rm BNS}}}=110-3840{\rm Gpc^{-3}yr^{-1}}$;
and the upper limit of BH-NS merger is $\mathscr{R_{{\rm BHNS}}}<600{\rm Gpc^{-3}yr^{-1}}$.

Three main evolutionary channels were proposed in the context of GW
mergers. The first deals with merger in dense environments such as
galactic centers or globular clusters \citep[e.g.][]{Rodriguez2016,Rodriguez2018,Leigh2018},
where binary mergers are catalyzed by strong interactions with stars
in these dense environment. In such environments, strong three-body
interactions lead to harden compact binaries (drive them to shorter
periods) and excite their eccentricities. Such models predict GW-production
rates in the range of $\sim2-20{\rm Gpc^{-3}yr^{-1}}.$ The second
evolutionary channel deals with the isolated evolution of initially
massive close binary stars \citep[e.g.][]{Belczynski2008,Belczynski2016,Dominik2012,Dominik2015,Mandel2016}.
Some of the massive close binaries strongly interact through one or
two common envelope phases \citep[e.g.][]{Dominik2012} in which the
interaction of a star with the envelope of an evolved companion leads
to its inspiral in the envelope and the production of a short period
binary. A fraction of the post-CE binaries are sufficiently close
to merge via GW emission within Hubble time. The large uncertainties
in the initial conditions of the binaries, the evolution in the common-envelope
phase, the natal-kick experienced by NS/BHs at birth; and the mass-loss
processes of massive stars give rise to a wide range of expected GW-sources
production rates in the range $\sim10^{-2}-10^{3}{\rm Gpc^{-3}yr^{-1}}.$
The third evolutionary channel deals with mergers induced by secular
evolution of triple systems either in the field \citep{Antonini2017}
or in dense environments \citep[e.g.][]{Samsing2018,Samsing2018a,Petrovich2017,Antonini2018,Antonini2012}.
In this channel the secular perturbations by a third companion (Lidov-Kozai
evolution \citep{Lidov1962,Kozai1962} ) can drive BBHs into high
eccentricities such that they merge within a Hubble-time; the rates
expected in this channel are $\sim0.5-15{\rm Gpc^{-3}yr^{-1}.}$

Here we present a fourth channel of binary evolution, in which we
focus of wide (SMA >$1000{\rm AU})$ BBHs in the field perturbed by
random fly-by interactions of field stars in their host galaxy. \citet{Kaib2014}
and \citet{Michaely2016} showed that although evolution of stars
and binaries in the low-density environment in the field is typically
thought to be collisionless, wide binary systems can be significantly
affected by fly-by interactions of field stars stars, and effectively
experience a collisional evolution. In particular, \citet{Kaib2014}
calculated the probability of a head-on collision between two main-sequence
stars in the Milky-Way Galaxy due to interaction with random stellar
perturbers. \citet{Michaely2016} followed these directions and suggested
a novel formation scenario for low-mass X-ray binaries from wide-binaries
in the field.

Here we show that collisional evolution in the field could also be
highly important for the formation of GW-sources. We analyze the evolution
of wide-orbit BBHs and show that a fraction of these can be driven
into high eccentricities and close pericenter distances due to interaction
with stellar perturbers. In cases where the pericenter distance of
a given binary is driven into a sufficiently small distance GW-emission
becomes significant and the binary rapidly loses angular momentum
and energy due to GW emission, and eventually inspirals and merges
as a GW-sources detectable by aLIGO/VIRGO.

This paper is organized as follows. In section \ref{sec:Analytic-model}
we present the analytic model, the basic assumptions and the calculations.
In section \ref{sec:Numerical-calculation} we present the numerical
verification to the analytic model. We discuss the results and summarize
in section \ref{sec:Discussio}. The numerical procedure, the equations
we integrate and the data analysis are described at length at appendix
\ref{sec:Appendix-I} and \ref{sec:Appendix-II}.

\section{\label{sec:Analytic-model}Analytic model}

\subsection{Formation Scenario \label{subsec:Basic-Formation-Scenario}}

We consider a wide BBH with semi-major axis (SMA) $a>10^{3}{\rm AU}$.
The binary resides in the field of the host galaxy and therefore be
affected by short duration dynamical interactions with field stars.
The dynamical encounters can typically be modeled through the impulse
approximation, i.e. in the regime where the interaction timescale
$t_{{\rm int}}\equiv b/v_{{\rm enc}}$(where $b$ is the closest approach
to the binary and $v_{{\rm enc}}$ is the velocity of the perturbing
mass) is much shorter than the BBH orbital period time, $P$. These
perturbations can torque the system and exchange orbital energy thereby
decreasing/increasing the binary semi-major axis $a$ and the binary
eccentricity $e$. If these interactions drive the system to a sufficiently
small pericenter passage, $q$, then the system can merge via GW emission
within Hubble time.

There are four relevant timescales for this impulsive treatment: the
interaction timescale $t_{{\rm int}}\equiv b/v_{{\rm enc}}$; the
binary orbital period $P$; the merger time from a specific binary
configuration via GW emission $T$; and the time between two consecutive
encounters of the system and a fly-by perturber, $t_{{\rm enc}}=1/f=\left(n_{*}\sigma v_{{\rm enc}}\right)^{-1}$
where $n_{*}$ is the stellar number density, $\sigma$ is the geometric
cross-section of the binary and the stellar fly-by.

We restrict our model to the impulsive regime, namely $t_{{\rm int}}\ll P$.
This gives upper bound to the closest approach distance $b$ and hence
limits the average time between encounters. For example, for a BBH
with a total mass of $20M_{\odot}$, SMA of $a\sim10^{4}{\rm AU}$
(hence $P\approx3\times10^{5}{\rm yr}$) and a typical velocity encounter
of $v_{{\rm enc}}=50{\rm kms^{-1}}$(velocity dispersion in the field)
we can restrict $b$ such that $t_{{\rm int}}=P/10$ $\left(P/100\right)$.
Hence we get $b=t_{{\rm int}}\times v_{{\rm enc}}=3\times10^{5}{\rm AU}\;\left(3\times10^{4}{\rm AU}\right)$.
Farther out flybys can also perturb the system and further excite
the system, but at very large separations the interaction become adiabatic
and the effects become small. We neglect the intermediate regime in
which the perturbation time and the orbital times are comparable,
which are likely to somewhat enhance the perturbation rates explored
here.

\subsection{\label{subsec:Analytic-Description}Analytic Description}

We consider the evolution of an ensemble of wide BBH binaries with
initial separations $a>10^{3}{\rm AU}$ and comparable component masses
$m_{1}\sim m_{2}=m_{{\rm BH}}$. For simplicity we assume all binaries
to have the same SMA, and a thermal distribution of orbital eccentricities,
$f(e)de=2ede$. In the following we derive the fraction of merging
systems within this ensemble and find its dependence on the SMA of
the binaries, $a$ and the environmental conditions, namely the stellar
density $n_{*}$ and velocity dispersion $\sigma_{v}=v_{{\rm enc}}$.

The timescale for a GW-merger of an isolated binary is given by \citet{Pet64}

\begin{equation}
t_{{\rm merger}}\approx\frac{a^{4}}{\beta'}\times\left(1-e^{2}\right)^{7/2}\label{eq:t_merger}
\end{equation}
with 
\[
\beta'=\frac{85}{3}\frac{G^{3}m_{1}m_{2}\left(m_{1}+m_{2}\right)}{c^{5}},
\]
where $G$ is Newton's constant and $c$ is the speed of light and
$m_{1}=m_{2}=m_{{\rm BH}}$. Given a binary with SMA $a$ we can solve
equation (\ref{eq:t_merger}) for the critical eccentricity $e_{c}$
required for the binary to merger within some merger time ${\rm T}=t_{{\rm merge}}$
\[
e_{c}=\left[1-\left(\frac{\beta'T}{a^{4}}\right)^{2/7}\right]^{1/2}.
\]
All systems with eccentricities equal or greater than $e_{c}$ would
therefore merge within this time-frame $T$. Hence, given a thermal
distribution of eccentricities we find the fraction of system that
merge within a time $T$ and lost from the ensemble to be: 
\begin{equation}
F_{q}=\int_{e_{c}}^{1}2ede=1-e_{c}^{2}=\left(\frac{\beta'T}{a^{4}}\right)^{2/7}.\label{eq:F_q}
\end{equation}
Following previous studies we term this ``loss'' region the ``loss-cone''
(see e.g. \citep{Hills1981}); after time $T$ all binaries in the
loss-cone merge via GW-emission and this phase-space region become
empty. However, binaries outside the loss-cone which do not merge
within this timescale, can be perturbed by a flyby encounter as to
change their angular momentum, and thereby enter and replenish the
loss cone. The average size of the phase-space region into which stars
are perturbed during a single orbital period is termed the smear cone,
defined by 
\begin{equation}
\theta=\frac{\left\langle \Delta v\right\rangle }{v_{k}}\label{eq:theta-1}
\end{equation}
where $v_{k}$ is the Keplerian velocity of the binary. The value
of $v_{k}$ can be calculated given that the average separation of
a Keplerian orbit is $\left\langle r\right\rangle =a\left(1+1/2e^{2}\right)$
and we approximate $e\rightarrow1$, namely 
\[
v_{k}=\left(\frac{Gm_{b}}{3a}\right)^{1/2}.
\]

$\left\langle \Delta v\right\rangle $ is the average change in the
velocity over an orbital period due to perturbations \citep{Hills1981}.
Let us consider fly-by interactions using the impulse approximation.
\citet{Hills1981} showed that on average the velocity change (for
a binary with SMA, $a$), to the binary components is of the order
of 
\begin{equation}
\left\langle \Delta v\right\rangle \simeq\frac{3Gam_{p}}{v_{{\rm enc}}b^{2}}\label{eq:Delta_V}
\end{equation}
where $v_{{\rm enc}}$ is the velocity of the fly-by star with respect
to the binary center of mass, $m_{p}$ is the perturber mass and $b$
is the closest approach distance of a fly-by. Therefore, the square
of the angular size of the smear cone cause by the impulse of the
fly-by on the binary is 
\begin{equation}
\theta^{2}=\frac{9G^{2}a^{2}m_{p}^{2}}{\left(v_{{\rm enc}}b^{2}\right)^{2}}\frac{3a}{Gm_{b}}=\frac{27Ga^{3}m_{p}^{2}}{m_{b}\left(v_{{\rm enc}}b^{2}\right)^{2}}\label{eq:theta_calc}
\end{equation}
and for $\theta\ll1$ we get the fractional size of the smear-cone
velocity space over the $4\pi$ sphere to be after a single passage
of the perturber 
\begin{equation}
F_{s}=\frac{\pi\theta^{2}}{4\pi}=\frac{27}{4}\left(\frac{m_{p}}{m_{b}}\right)^{2}\left(\frac{Gm_{b}}{av_{{\rm enc}}^{2}}\right)\left(\frac{a}{b}\right)^{4}.\label{eq:F_s}
\end{equation}
It is evident from (\ref{eq:F_s}) that for a given binary the size
of the smear cone depends on the perturber quantities, i.e. mass,
velocity and the closest approach. The ratio of $F_{s}$ to $F_{q}$
indicates the fraction of the loss cone filled after a single fly-by.
\begin{equation}
\frac{F_{s}}{F_{q}}=\frac{27}{4}\left(\frac{m_{p}}{m_{b}}\right)^{2}\left(\frac{Gm_{b}}{av_{{\rm enc}}^{2}}\right)\left(\frac{a}{b}\right)^{4}\left(\frac{a^{4}}{\beta'T}\right)^{2/7}\label{eq:Fs/Fq}
\end{equation}

In the case where the loss cone is continuously full ($F_{s}=F_{q}$)
the depletion rate only depends on the loss cone size, $F_{q}$ and
the merger time , $T$. Hence the loss rate for the full lose cone
is given by: 
\begin{equation}
\dot{L}_{q}=\frac{F_{q}}{T}.\label{eq:Loss_rate-1}
\end{equation}
Note that the loss rate is independent of the stellar density in the
field, i.e. once the stellar density is sufficiently large as to fill
the loss-cone, the loss rate is saturated, and becomes independent
of the perturbation rate. Furthermore, one can see from Eq. (\ref{eq:Loss_rate-1})
that the full loss-cone rate scales like $\dot{L}\propto F_{q}\propto a^{-8/7}$,
i.e. the full loss-cone rate decreases with increasing SMA.

On the other hand, tighter binaries are less susceptible for change
due to a fly-by, this is evident from equation (\ref{eq:F_s}). Therefore
closer from a critical SMA we expect that the loss cone will not be
full all the time, in this ``empty loss cone'' case the loss rate
depends on the rate of orbits being kicked into the loss cone: 
\begin{equation}
f=n_{*}\sigma v_{{\rm enc}}.\label{eq:f fly_by rate}
\end{equation}
where $n_{*}$ is the stellar density, $\sigma=\pi b^{2}$ is the
geometric cross-section.

The condition for the loss cone to be continuously full is that the
loss-cone orbits are replenished at least as fast as they are depleted
due to the GW emission. This occurs when the rate of fly-by's that
enter orbits to the loss cone,$f$ is equal to the rate of which orbits
are depleted from the loss cone, $1/T$: 
\begin{equation}
n_{*}\pi b^{2}v_{{\rm enc}}=\frac{1}{T}.\label{eq:condition_equilibirum}
\end{equation}

Furthermore, the condition for the loss cone to be continuously full
is that the sizes of the lose-cone and the smear-cone are equal. This
equilibrium occurs when 
\begin{equation}
\frac{F_{s}}{F_{q}}=\frac{27}{4}\left(\frac{m_{p}}{m_{b}}\right)^{2}\left(\frac{Gm_{b}}{av_{{\rm enc}}^{2}}\right)\left(\frac{a}{b}\right)^{4}\left(\frac{a^{4}}{\beta'T}\right)^{2/7}=1.\label{eq:F_s/F_q}
\end{equation}
A stellar fly-by is sufficiently strong as to replenish the loss cone
if 
\begin{equation}
\left(v_{{\rm enc}}b^{2}\right)^{2}\leq\frac{27}{4}\frac{Gm_{p}^{2}a^{29/7}}{m_{b}\left(\beta'T\right)^{2/7}}
\end{equation}
Plugging this to equation (\ref{eq:condition_equilibirum}) we get
an equation for the critical SMA that separates the empty and full
loss-cone regimes: 
\begin{equation}
a_{{\rm crit}}=\left[\frac{4}{27}\frac{m_{b}\beta'^{2/7}T^{-12/7}}{Gm_{p}^{2}n_{*}^{2}\pi^{2}}\right]^{7/29}.
\end{equation}

Using the critical SMA we can calculate the merger probability in
each of these regimes, $a<a_{{\rm crit}}$ (empty) and $a>a_{{\rm crit}}$
(full). We denote $F_{q}$ as the fraction of wide binaries destroyed
after time $T$, and therefore $\left(1-F_{q}\right)$ represents
the fraction of binaries that survive as wide binaries at the relevant
timescale. For the empty loss-cone regime ($a<a_{{\rm crit}})$ the
relevant timescale is $1/f$; for the full loss-cone regime ($a>a_{{\rm crit}})$
the relevant timescale is $T$. Therefore, $\left(1-F_{q}\right)$
is a monotonically decreasing function of time, and the probability
for a merger of a wide binary is 
\begin{equation}
L_{a<a_{{\rm crit}}}=1-\left(1-F_{q}\right)^{t\cdot f}\label{eq:empty_Prob}
\end{equation}
where $t$ is the time since birth of the binary. As one can expect
the probability only depends on the size of the loss cone and the
rate of interactions. For the limit of $t\cdot f\cdot F_{q}\ll1$
we can expand equation (\ref{eq:empty_Prob}) and take the leading
term, to find the loss probability to be approximated by 
\begin{equation}
L_{a<a_{{\rm crit}}}=tfF_{q}.\label{eq:empty_prop_approx}
\end{equation}
given that 
\begin{equation}
f=n_{*}\pi\left(\frac{27}{4}\frac{Gm_{p}^{2}a^{29/7}}{m_{b}\left(\beta'T\right)^{2/7}}\right)^{1/2}
\end{equation}
together with equation (\ref{eq:F_q}) we get: 
\begin{equation}
L_{a<a_{{\rm crit}}}=tn_{*}m_{*}a^{13/14}\times\left(\frac{27G\left(\beta'T\right)^{2/7}}{4m_{b}}\right)^{1/2}\label{eq:empty_propb_express}
\end{equation}

In the full loss-cone regime the limiting factor is not the value
of $f$, but the merger timescale $T$. Therefore, the full expression
for the loss probability for $a>a_{{\rm crit}}$ is 
\begin{equation}
L_{a>a_{{\rm crit}}}=1-\left(1-F_{q}\right)^{t/T}.\label{eq:full_prob}
\end{equation}
In the limit of $F_{q}\cdot t/T\ll1$ we can approximate the probability
by 
\begin{equation}
L_{a>a_{{\rm crit}}}=tF_{q}\frac{1}{T}=t\left(\frac{\beta'T}{a^{4}}\right)^{2/7}\frac{1}{T}=ta^{-8/7}\times T^{-5/7}\beta'^{2/7}.\label{eq:full_prob_apx_express}
\end{equation}

\begin{figure}
\includegraphics[width=0.9\columnwidth]{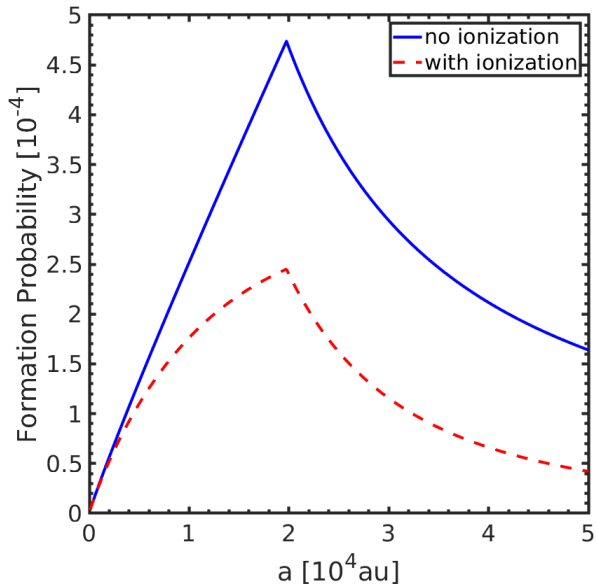} \caption{\label{fig:The-merger-probability}The merger probability of BBH with
$m_{b}=20M_{\odot}$ with fly-by mass $m_{p}=0.6M_{\odot}$ and $v_{{\rm enc}}=50{\rm kms^{-1}}$.
The stellar density number is $n_{*}=0.1{\rm pc^{-3}}$. The probability
it calculated after $t=10{\rm Gyr}$ since the BBH was formed. The
peak probability is achieved at $a=a_{{\rm crit}}$. The merger time
$T=1{\rm Myr}$, this value is chosen to ensure that the binary will
merge between two consecutive encounters with stellar fly-bys. Blue
solid line without accounting for ionization; red dashed line accounting
for ionization.}
\end{figure}

The above treatment neglects the fact that perturbations may also
``ionize'' a binary and destroy it, namely, the binary is disrupted
by the random fly-bys. Such ionization process decreases the available
number of wide binaries. To account for the ionization process we
consider the finite lifetime of wide binaries due to fly-bys using
the approximate relation given by \citep{Bahcall1985} for $t_{1/2}$,
the half-life time of a wide binary evolving through encounters 
\begin{equation}
t_{1/2}=0.00233\frac{v_{{\rm enc}}}{Gm_{p}n_{*}a}.\label{eq:t_half_life}
\end{equation}
Taking this into account we can correct for eq. (\ref{eq:empty_propb_express})
and eq. (\ref{eq:full_prob_apx_express}) to get 
\begin{equation}
L_{a<a_{{\rm crit}}}=\tau n_{*}m_{*}a^{13/14}\times\left(\frac{27G\left(\beta'T\right)^{2/7}}{4m_{b}}\right)^{1/2}\left(1-e^{-t/\tau}\right)\label{eq:ionization empty}
\end{equation}
and 
\begin{equation}
L_{a>a_{{\rm crit}}}=\tau a^{-8/7}\times T^{-5/7}\beta'^{2/7}\left(1-e^{-t/\tau}\right),\label{eq:ionization full}
\end{equation}
where $\tau=t_{1/2}/\ln2$ is the mean-lifetime of the binary.

In order to estimate the number of systems observable within a year
in aLIGO we first calculate the number of systems merging in a Milky-Way
(MW)-like galaxy per unit time. In order to do that we need to integrate
over all SMA in a given stellar density and over all stellar densities
in the galaxy we model. We follow a similar calculation from \citet{Michaely2016}.
We model the Galaxy in the following way, let $dN\left(r\right)=n_{*}\left(r\right)\cdot2\pi\cdot r\cdot h\cdot dr$
be the the number of stars in a region $dr$ (and scale height $h$),
located at distance $r$ from the center of the Galaxy. Following
\citep{Kaib2014} and references within we model the Galactic stellar
density in the Galactic disk as follows 
\begin{equation}
n_{*}\left(r\right)=n_{0}e^{\left(-\left(r-r_{\odot}\right)/R_{l}\right)},
\end{equation}
where $n_{0}=0.1{\rm pc^{-3}}$ is the stellar density near our sun,
$R_{l}=2.6{\rm kpc}$ \citep{Juric2008} is the galactic length scale
and $r_{\odot}=8{\rm kpc}$ is the distance of the sun from the galactic
center. Integrating over the stellar densities throughout the Galaxy
we can obtain the total number of mergers through this process. Next
we account for the fraction of wide BBH systems from the entire population
of starts in the Galaxy. We use the following standard values. Given
a Kropa initial mass function \citep{Kroupa2001}, the fraction of
the stars that evolve to become BHs is $f_{{\rm primary}}\approx10^{-3}$.
If we assume most BHs form without any natal-kick ( similar assumptions
were taken in other works \citep[e.g.][]{Mandel2016a,Belczynski2016}),
we can expect all BHs to be in binary (or higher multiplicity) systems
and the fraction $f_{{\rm bin}}=1$, consistent with the binary fraction
inferred for the O-stars progenitors of BHs \citep{Moe2016,Duchene2013,Sana2014}.
Next we assume a uniform distribution of the mass ratios, $Q\in\left(0.1,1\right)$
\citep{Moe2016,Duchene2013}, to get a fraction of secondaries that
evolve into BHs of $f_{{\rm secondary}}\approx0.4$. We also assume
that the SMA has a log-uniform distribution (Opik law) and therefore
the fraction of systems with SMA larger than $10^{3}{\rm AU}$ is
$f_{{\rm wide}}\approx f_{{\rm bin}}\times0.2$. This value is actually
a lower limit for massive binaries, recently \citet{Igoshev2019}
found that a wide binary fraction for massive B-stars to be $f_{{\rm wide}}\approx0.5$,
and theoretical models suggest that the fraction of wide binary O-stars
and BHs could be close to unity \citep{Perets2012a}, $f_{{\rm wide}}=1$,
and we therefore expect a wide-binary fraction in the range $0.2-1$,
in the following we use $f_{{\rm wide}}=0.5$. 
\[
f_{{\rm BBH}}\approx f_{{\rm primary}}\times f_{{\rm secondary}}\times f_{{\rm wide}}\approx
\]
\begin{equation}
2\times10^{-4}\left(\frac{f_{{\rm primary}}}{10^{-3}}\right)\left(\frac{f_{{\rm seoncdary}}}{0.4}\right)\left(\frac{f_{{\rm wide}}}{0.5}\right).
\end{equation}
The number of merging BBH per ${\rm Myr}$ from this channel for a
MW-like Galaxy is 
\begin{equation}
\Gamma=\int\int L\left(a,r\right)\times f_{{\rm BBH}}{\rm \times\left(\frac{1Myr}{10{\rm Gyr}}\right)dadr\approx0.42Myr^{-1}}.
\end{equation}

Following \citet{Belczynski2016} we calculate the merger rate, $\mathscr{R}$
per ${\rm Gpc^{3}}$ by using the following estimate 
\[
\mathscr{R}=10^{3}\rho_{{\rm gal}}\times\Gamma\approx
\]
\begin{equation}
4.9\left(\frac{f_{{\rm primary}}}{10^{-3}}\right)\left(\frac{f_{{\rm seoncdary}}}{0.4}\right)\left(\frac{f_{{\rm wide}}}{0.5}\right){\rm Gpc^{3}yr^{-1}}
\end{equation}
while $\rho_{{\rm gal}}$ is local density of the MW-like galaxies
with the value of $\rho_{{\rm gal}}=0.0116{\rm Mpc^{-3}}$(e.g. \citet{Kopparapu2008})
and $\Gamma$ is given in the units of ${\rm Myr^{-1}.}$

\section{\label{sec:Numerical-calculation}Numerical calculation}

In this section we describe the numerical calculation we preform.
We simulate the evolution of a binary BH with masses of $m_{1}=m_{2}=10M_{\odot}$
for $10{\rm Gyr}.$ We treat the evolution by considering both the
evolution of the binary between encounters, and in-particular the
effects of GW-emission, as well as the change of the binary orbital
elements due to the impulsive fly-by encounters with field stars.

In order to calculate the average time between encounters we use the
rate $f=n_{*}\sigma\left\langle v_{{\rm enc}}\right\rangle $, where
$n_{*}$ is the stellar number density, taken to be the solar neighborhood
value of $n_{*}=0.1{\rm pc^{-3}}$; $\left\langle v_{{\rm enc}}\right\rangle $
is the velocity of the perturber as measured from the binary center
of mass, where we set $\left\langle v_{{\rm enc}}\right\rangle =50{\rm kms^{-1}}$
similar to the velocity dispersion in the solar neighborhood; and
$\sigma$ is the interaction cross-section. We focus on the impulsive
regime, namely $t_{{\rm int}}\ll P$ (see subsection \ref{subsec:Basic-Formation-Scenario}).
With these values the largest closest approach distance $b$ for which
an encounter can be considered as impulsive is \textbf{$b_{{\rm max}}=5\times10^{4}{\rm AU}.$
}The average time between such impulsive encounters is given by $t_{{\rm enc}}=1/f\approx1{\rm Myr}$.
Therefore we randomly sample the time between encounters from an exponential
distribution with a mean $f$ (due to the Poisson distribution of
encounter times).

We initialize the wide binary with a SMA $a$ and eccentricity $e$.
At each step we first find the next encounter time $t_{{\rm enc}}$,
and evolve the binary for $t_{{\rm enc}}$ through the equations of
motion given by \citet{Pet64}, 
\begin{equation}
\frac{da}{dt}=-\frac{64}{5}\frac{G^{3}m_{1}m_{2}\left(m_{1}+m_{2}\right)}{c^{5}a^{3}\left(1-e^{2}\right)^{7/2}}\left(1+\frac{73}{24}e^{2}+\frac{37}{96}e^{4}\right)
\end{equation}
\begin{equation}
\frac{de}{dt}=-e\frac{304}{15}\frac{G^{3}m_{1}m_{2}\left(m_{1}+m_{2}\right)}{c^{5}a^{4}\left(1-e^{2}\right)^{5/2}}\left(1+\frac{121}{304}e^{2}\right)
\end{equation}
where $G$ is Newton's constant and $c$ is the speed of light. If
the binary did not merge through GW-emission by the time of the next
encounter we simulate the impulsive interaction with a perturber with
velocity of $v_{{\rm enc}}$ drawn from a Maxwellian distribution
with velocity dispersion $\left\langle v_{{\rm enc}}\right\rangle $
and a mass of $m_{p}=0.6M_{\odot}$, typical for stars in the field.
After changing the binary orbital parameters due to the encounter
we continue to evolve the binary until the next encounter and so on,
until the binary merges, disrupts or the maximal simulation time of
10 Gyrs is reached. 

\begin{figure}
\includegraphics[width=0.9\columnwidth]{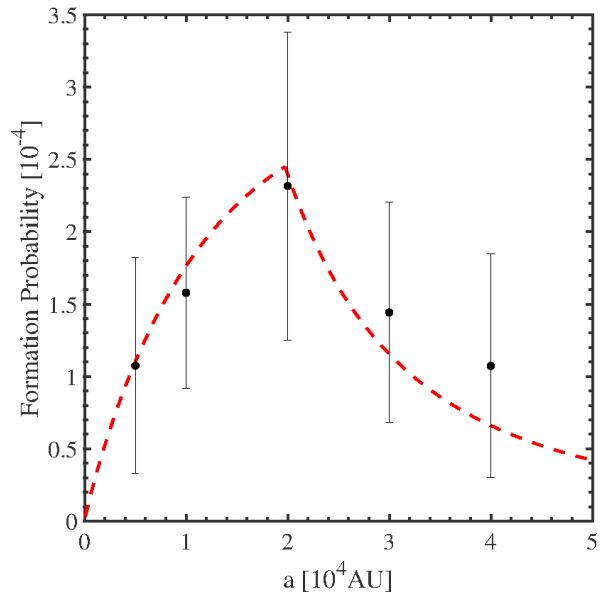}\caption{\label{fig:Numerical-verification-of}Numerical verification of the
analytic model. Red dashed line (same as Fig. (\ref{fig:The-merger-probability}))
is the theoretical probability for a merger as a function of initial
SMA. The black circles are the estimated probabilities from the numerical
simulation (see section (\ref{sec:Numerical-calculation}) for details).
The error-bar represent one standard deviation from the estimated
value. }
\end{figure}

The numerical results are presented in Figure \ref{fig:Numerical-verification-of};
the numerical result are highly consistent with the result of the
analytic model. A more detailed technical description of the numerical
procedure, equations and analysis is given in the appendix \ref{sec:Appendix-I}
and \ref{sec:Appendix-II}.

\section{\label{sec:Discussio}Discussion and summary}

In this paper we explore a novel channel for the production of BBH
GW-sources from wide ($>10^{3}$AU) binaries in the field. Such binaries
are sensitive to perturbations by stellar-fly-bys even in the low
density environment in the field. We find that a fraction of all wide-binaries
attain sufficiently close-approaches (as their orbits are excited
to very high eccentricities) as to inspiral and eventually merge through
GW-emmision before any consecutive encounter can change the orbit.

The merger rate strongly depends on the natal-kicks given to BHs at
birth, which are poorly constrained \citep[e.g.][]{Repetto2012,Repetto2017}.
In particular, it is still unknown whether a BH receives a momentum
kick at birth like a NS, or forms without any natal-kick following
a failed supernova or a large amount of fallback \citep[e.g.][]{Ertl2015,Belczynski2004,Belczynski2008}.
Previous models that were able to produce rates comparable to the
rate inferred from observations had typically taken similar assumptions
of zero kick velocities (for all BHs, or at least for all BHs more
massive than 10 ${\rm M}_{\odot}$), while models assuming higher
natal kicks produced significantly lower rates \citep{Belczynski2008,Dominik2015}.
In our case low natal-kicks can unbind the wide binaries, lowering
their fraction. In principle, in models where wide-binaries form following
the dispersal of their birth-cluster on longer time-scales \citep{Kouwenhoven2010,Perets2012a},
BH may acquire wide companions well after their formation. Nevertheless,
even in these cases the BHs need to be retained in the cluster until
its dispersal, and therefore the natal kick needs to be sufficiently
low for a BH not to escape the cluster. We conclude that adapting
similar no-kick assumptions for BHs (as done by other potentially
successful scenarios) suggests the wide-binary channel explored here
can give rise to a high production rate of GW-sources from BBH mergers
of perturbed ultra-wide binaries.

Beside the rate estimate, $\sim10\times f_{{\rm wide}}{\rm \,yr^{-1}}{\rm Gpc^{-3}}$,
our proposed evolutionary channel gives rise to specific characteristics
of the BBH mergers, which together can provide a distinct signature
for this channel, as we discuss in the following.

Eqs. (\ref{eq:ionization empty}) and (\ref{eq:ionization full})
describe the probability dependency for a given environment, namely
the stellar density $n_{*}$ and the encounter velocity $v_{{\rm enc}}.$
We note the in both equations there is a $\tau\left(1-e^{-t/\tau}\right)$
dependency. Hence following eq. (\ref{eq:t_half_life}) the merger
probability increases with the encounter velocity. For example, taking
the same environment as assumed in section \ref{subsec:Analytic-Description}
but with $v_{{\rm enc}}=200{\rm kms^{-1}}$ gives a factor of $\sim1.92$
higher rate of BBH GW-sources. We therefore expect a preference for
host galaxies with higher velocity dispersion.

This model is sensitive to extreme mass ratio. The equations that
govern the rates depend on $\beta'$. When the binary mass is kept
constant but the mass ratio $Q$ is varied we get the following dependence
on $Q$ 
\begin{equation}
\beta'=\frac{85}{3}\frac{G^{3}m_{1}m_{2}\left(m_{1}+m_{2}\right)}{c^{5}}\propto\frac{Q}{\left(1+Q\right)^{2}},
\end{equation}
and since $Q$ is a monotonically increasing function, equal mass
components have the highest probability to merge. Moreover, the merger
rate also has a monotonic dependence on the total binary mass, due
to the complex mass dependence in the loss-cone analysis and the effects
of ionization (see Eqs. \ref{eq:ionization empty} and \ref{eq:ionization full}).
Hence overall we expect a preference towards GW-sources from more
massive binaries and higher mass-ratios. Furthermore, in this channel
the spins of the BHs are likely uncorrelated given the origin of the
BH components from very wide separations (or a random capture) and
we therefore expect the spins of the merging BBH components to be
randomly (mis-)alligned, in contrast with e.g. the isolated binary
evolution channel; \citep[e.g.]{Mandel2016}. Given the long time-scale
for inspiral from large separations we also expect BBHs to fully circularize
by the time they reach the aLIGO band and to not produce any eccentric
binaries at these frequencies, in contrast with some of the dynamical
channels. Finally, unlike the isolated binary channel, which predicts
a delay time dependence of $\propto t^{-1}$ \citep{Dominik2015}
our model, which have no time dependency on the merger probability,
generally predicts a uniform delay time distribution.

In summary, the wide-binary origin for BBH GW mergers can give rise
to a potential rate of $\sim10\times f_{{\rm wide}}$ ${\rm yr}^{-1}$${\rm Gpc^{-3}}$
(where $f_{{\rm wide}}$ can plausibly reside in the range $0.2-1$),
comparable to (the lower range of) the observationally inferred rate
of $\sim10-110$ ${\rm yr}^{-1}$${\rm Gpc^{-3}}$ from aLIGO/VIRGO
detection, and is strongly dependent on the natal-kicks imparted to
BHs at birth. It can be characterized by the following signatures:
(1) A slight preference for high mass ratio BBH GW-sources. (2) A
preference for more massive BBH. (3) Typically randomly misaligned
spin-orbits BHs. (4) Circular orbits in the aLIGO band. (5) Preference
for high velocity dispersion host galaxies/environments. (6) A uniform
delay-time distribution. 
We acknowledge support from the ISF-ICORE grant 1829/12. The authors
acknowledge the University of Maryland supercomputing resources (http://hpcc.umd.edu)
made available for conducting the research reported in this paper.
EM would like to thank Coleman Miller and Johan Samsing for stimulating
discussions regarding this work. 
\bibliographystyle{plainnat}

\begin{thebibliography}{43}
\providecommand{\natexlab}[1]{#1}
\providecommand{\url}[1]{\texttt{#1}}
\expandafter\ifx\csname urlstyle\endcsname\relax
  \providecommand{\doi}[1]{doi: #1}\else
  \providecommand{\doi}{doi: \begingroup \urlstyle{rm}\Url}\fi

\bibitem[{Antognini} et~al.(2014){Antognini}, {Shappee}, {Thompson}, and
  {Amaro-Seoane}]{Antognini2014}
J.~M. {Antognini}, B.~J. {Shappee}, T.~A. {Thompson}, and P.~{Amaro-Seoane}.
\newblock {Rapid eccentricity oscillations and the mergers of compact objects
  in hierarchical triples}.
\newblock \emph{\mnras}, 439:\penalty0 1079--1091, March 2014.
\newblock \doi{10.1093/mnras/stu039}.

\bibitem[{Antonini} and {Perets}(2012)]{Antonini2012}
F.~{Antonini} and H.~B. {Perets}.
\newblock {Secular Evolution of Compact Binaries near Massive Black Holes:
  Gravitational Wave Sources and Other Exotica}.
\newblock \emph{\apj}, 757:\penalty0 27, September 2012.
\newblock \doi{10.1088/0004-637X/757/1/27}.

\bibitem[{Antonini} et~al.(2014){Antonini}, {Murray}, and
  {Mikkola}]{Antonini2014}
F.~{Antonini}, N.~{Murray}, and S.~{Mikkola}.
\newblock {Black Hole Triple Dynamics: A Breakdown of the Orbit Average
  Approximation and Implications for Gravitational Wave Detections}.
\newblock \emph{\apj}, 781:\penalty0 45, January 2014.
\newblock \doi{10.1088/0004-637X/781/1/45}.

\bibitem[{Antonini} et~al.(2017){Antonini}, {Toonen}, and
  {Hamers}]{Antonini2017}
F.~{Antonini}, S.~{Toonen}, and A.~S. {Hamers}.
\newblock {Binary Black Hole Mergers from Field Triples: Properties, Rates, and
  the Impact of Stellar Evolution}.
\newblock \emph{\apj}, 841:\penalty0 77, June 2017.
\newblock \doi{10.3847/1538-4357/aa6f5e}.

\bibitem[{Antonini} et~al.(2018){Antonini}, {Rodriguez}, {Petrovich}, and
  {Fischer}]{Antonini2018}
F.~{Antonini}, C.~L. {Rodriguez}, C.~{Petrovich}, and C.~L. {Fischer}.
\newblock {Precessional dynamics of black hole triples: binary mergers with
  near-zero effective spin}.
\newblock \emph{\mnras}, 480:\penalty0 L58--L62, October 2018.
\newblock \doi{10.1093/mnrasl/sly126}.

\bibitem[{Bahcall} et~al.(1985){Bahcall}, {Hut}, and {Tremaine}]{Bahcall1985}
J.~N. {Bahcall}, P.~{Hut}, and S.~{Tremaine}.
\newblock {Maximum mass of objects that constitute unseen disk material}.
\newblock \emph{\apj}, 290:\penalty0 15--20, March 1985.
\newblock \doi{10.1086/162953}.

\bibitem[{Belczynski} et~al.(2002){Belczynski}, {Kalogera}, and
  {Bulik}]{Belczynski2002}
K.~{Belczynski}, V.~{Kalogera}, and T.~{Bulik}.
\newblock {A Comprehensive Study of Binary Compact Objects as Gravitational
  Wave Sources: Evolutionary Channels, Rates, and Physical Properties}.
\newblock \emph{\apj}, 572:\penalty0 407--431, June 2002.
\newblock \doi{10.1086/340304}.

\bibitem[{Belczynski} et~al.(2004){Belczynski}, {Sadowski}, and
  {Rasio}]{Belczynski2004}
K.~{Belczynski}, A.~{Sadowski}, and F.~A. {Rasio}.
\newblock {A Comprehensive Study of Young Black Hole Populations}.
\newblock \emph{\apj}, 611:\penalty0 1068--1079, August 2004.
\newblock \doi{10.1086/422191}.

\bibitem[{Belczynski} et~al.(2007){Belczynski}, {Taam}, {Kalogera}, {Rasio},
  and {Bulik}]{Belczynski2007}
K.~{Belczynski}, R.~E. {Taam}, V.~{Kalogera}, F.~A. {Rasio}, and T.~{Bulik}.
\newblock {On the Rarity of Double Black Hole Binaries: Consequences for
  Gravitational Wave Detection}.
\newblock \emph{\apj}, 662:\penalty0 504--511, June 2007.
\newblock \doi{10.1086/513562}.

\bibitem[{Belczynski} et~al.(2008){Belczynski}, {Kalogera}, {Rasio}, {Taam},
  {Zezas}, {Bulik}, {Maccarone}, and {Ivanova}]{Belczynski2008}
K.~{Belczynski}, V.~{Kalogera}, F.~A. {Rasio}, R.~E. {Taam}, A.~{Zezas},
  T.~{Bulik}, T.~J. {Maccarone}, and N.~{Ivanova}.
\newblock {Compact Object Modeling with the StarTrack Population Synthesis
  Code}.
\newblock \emph{\apjs}, 174:\penalty0 223-260, January 2008.
\newblock \doi{10.1086/521026}.

\bibitem[{Belczynski} et~al.(2016){Belczynski}, {Repetto}, {Holz},
  {O'Shaughnessy}, {Bulik}, {Berti}, {Fryer}, and {Dominik}]{Belczynski2016}
K.~{Belczynski}, S.~{Repetto}, D.~E. {Holz}, R.~{O'Shaughnessy}, T.~{Bulik},
  E.~{Berti}, C.~{Fryer}, and M.~{Dominik}.
\newblock {Compact Binary Merger Rates: Comparison with LIGO/Virgo Upper
  Limits}.
\newblock \emph{\apj}, 819:\penalty0 108, March 2016.
\newblock \doi{10.3847/0004-637X/819/2/108}.

\bibitem[{Collins} and {Sari}(2008)]{Collins2008}
B.~F. {Collins} and R.~{Sari}.
\newblock {L{\'e}vy Flights of Binary Orbits due to Impulsive Encounters}.
\newblock \emph{\aj}, 136:\penalty0 2552--2562, December 2008.
\newblock \doi{10.1088/0004-6256/136/6/2552}.

\bibitem[{de Mink} and {Belczynski}(2015)]{deMink2015}
S.~E. {de Mink} and K.~{Belczynski}.
\newblock {Merger Rates of Double Neutron Stars and Stellar Origin Black Holes:
  The Impact of Initial Conditions on Binary Evolution Predictions}.
\newblock \emph{\apj}, 814:\penalty0 58, November 2015.
\newblock \doi{10.1088/0004-637X/814/1/58}.

\bibitem[{Dominik} et~al.(2012){Dominik}, {Belczynski}, {Fryer}, {Holz},
  {Berti}, {Bulik}, {Mandel}, and {O'Shaughnessy}]{Dominik2012}
M.~{Dominik}, K.~{Belczynski}, C.~{Fryer}, D.~E. {Holz}, E.~{Berti},
  T.~{Bulik}, I.~{Mandel}, and R.~{O'Shaughnessy}.
\newblock {Double Compact Objects. I. The Significance of the Common Envelope
  on Merger Rates}.
\newblock \emph{\apj}, 759:\penalty0 52, November 2012.
\newblock \doi{10.1088/0004-637X/759/1/52}.

\bibitem[{Dominik} et~al.(2015){Dominik}, {Berti}, {O'Shaughnessy}, {Mandel},
  {Belczynski}, {Fryer}, {Holz}, {Bulik}, and {Pannarale}]{Dominik2015}
M.~{Dominik}, E.~{Berti}, R.~{O'Shaughnessy}, I.~{Mandel}, K.~{Belczynski},
  C.~{Fryer}, D.~E. {Holz}, T.~{Bulik}, and F.~{Pannarale}.
\newblock {Double Compact Objects III: Gravitational-wave Detection Rates}.
\newblock \emph{\apj}, 806:\penalty0 263, June 2015.
\newblock \doi{10.1088/0004-637X/806/2/263}.

\bibitem[{Duch{\^e}ne} and {Kraus}(2013)]{Duchene2013}
G.~{Duch{\^e}ne} and A.~{Kraus}.
\newblock {Stellar Multiplicity}.
\newblock \emph{\araa}, 51:\penalty0 269--310, August 2013.
\newblock \doi{10.1146/annurev-astro-081710-102602}.

\bibitem[{Ertl} et~al.(2015){Ertl}, {Janka}, {Woosley}, {Sukhbold}, and
  {Ugliano}]{Ertl2015}
T.~{Ertl}, H.-T. {Janka}, S.~E. {Woosley}, T.~{Sukhbold}, and M.~{Ugliano}.
\newblock {A two-parameter criterion for classifying the explodability of
  massive stars by the neutrino-driven mechanism}.
\newblock \emph{ArXiv e-prints}, March 2015.

\bibitem[{Hills}(1981)]{Hills1981}
J.~G. {Hills}.
\newblock {Comet showers and the steady-state infall of comets from the Oort
  cloud}.
\newblock \emph{\aj}, 86:\penalty0 1730--1740, November 1981.
\newblock \doi{10.1086/113058}.

\bibitem[{Igoshev} and {Perets}(2019)]{Igoshev2019}
A.~P. {Igoshev} and H.~B. {Perets}.
\newblock {Wide binary companions to massive stars and their use in
  constraining natal kicks}.
\newblock \emph{arXiv e-prints}, January 2019.

\bibitem[{Juri{\'c}} et~al.(2008){Juri{\'c}}, {Ivezi{\'c}}, {Brooks}, {Lupton},
  {Schlegel}, {Finkbeiner}, {Padmanabhan}, {Bond}, {Sesar}, {Rockosi}, {Knapp},
  {Gunn}, {Sumi}, {Schneider}, {Barentine}, {Brewington}, {Brinkmann},
  {Fukugita}, {Harvanek}, {Kleinman}, {Krzesinski}, {Long}, {Neilsen}, {Nitta},
  {Snedden}, and {York}]{Juric2008}
M.~{Juri{\'c}}, {\v Z}.~{Ivezi{\'c}}, A.~{Brooks}, R.~H. {Lupton},
  D.~{Schlegel}, D.~{Finkbeiner}, N.~{Padmanabhan}, N.~{Bond}, B.~{Sesar},
  C.~M. {Rockosi}, G.~R. {Knapp}, J.~E. {Gunn}, T.~{Sumi}, D.~P. {Schneider},
  J.~C. {Barentine}, H.~J. {Brewington}, J.~{Brinkmann}, M.~{Fukugita},
  M.~{Harvanek}, S.~J. {Kleinman}, J.~{Krzesinski}, D.~{Long}, E.~H. {Neilsen},
  Jr., A.~{Nitta}, S.~A. {Snedden}, and D.~G. {York}.
\newblock {The Milky Way Tomography with SDSS. I. Stellar Number Density
  Distribution}.
\newblock \emph{\apj}, 673:\penalty0 864--914, February 2008.
\newblock \doi{10.1086/523619}.

\bibitem[{Kaib} and {Raymond}(2014)]{Kaib2014}
N.~A. {Kaib} and S.~N. {Raymond}.
\newblock {Very Wide Binary Stars as the Primary Source of Stellar Collisions
  in the Galaxy}.
\newblock \emph{\apj}, 782:\penalty0 60, February 2014.
\newblock \doi{10.1088/0004-637X/782/2/60}.

\bibitem[{Kopparapu} et~al.(2008){Kopparapu}, {Hanna}, {Kalogera},
  {O'Shaughnessy}, {Gonz{\'a}lez}, {Brady}, and {Fairhurst}]{Kopparapu2008}
R.~K. {Kopparapu}, C.~{Hanna}, V.~{Kalogera}, R.~{O'Shaughnessy},
  G.~{Gonz{\'a}lez}, P.~R. {Brady}, and S.~{Fairhurst}.
\newblock {Host Galaxies Catalog Used in LIGO Searches for Compact Binary
  Coalescence Events}.
\newblock \emph{\apj}, 675:\penalty0 1459--1467, March 2008.
\newblock \doi{10.1086/527348}.

\bibitem[{Kouwenhoven} et~al.(2010){Kouwenhoven}, {Goodwin}, {Parker},
  {Davies}, {Malmberg}, and {Kroupa}]{Kouwenhoven2010}
M.~B.~N. {Kouwenhoven}, S.~P. {Goodwin}, R.~J. {Parker}, M.~B. {Davies},
  D.~{Malmberg}, and P.~{Kroupa}.
\newblock {The formation of very wide binaries during the star cluster
  dissolution phase}.
\newblock \emph{\mnras}, 404:\penalty0 1835--1848, June 2010.
\newblock \doi{10.1111/j.1365-2966.2010.16399.x}.

\bibitem[{Kozai}(1962)]{Kozai1962}
Y.~{Kozai}.
\newblock {Secular perturbations of asteroids with high inclination and
  eccentricity}.
\newblock \emph{\aj}, 67:\penalty0 591, 1962.

\bibitem[{Kroupa}(2001)]{Kroupa2001}
P.~{Kroupa}.
\newblock {On the variation of the initial mass function}.
\newblock \emph{\mnras}, 322:\penalty0 231--246, April 2001.
\newblock \doi{10.1046/j.1365-8711.2001.04022.x}.

\bibitem[{Leigh} et~al.(2018){Leigh}, {Geller}, {McKernan}, {Ford}, {Mac Low},
  {Bellovary}, {Haiman}, {Lyra}, {Samsing}, {O'Dowd}, {Kocsis}, and
  {Endlich}]{Leigh2018}
N.~W.~C. {Leigh}, A.~M. {Geller}, B.~{McKernan}, K.~E.~S. {Ford}, M.-M. {Mac
  Low}, J.~{Bellovary}, Z.~{Haiman}, W.~{Lyra}, J.~{Samsing}, M.~{O'Dowd},
  B.~{Kocsis}, and S.~{Endlich}.
\newblock {On the rate of black hole binary mergers in galactic nuclei due to
  dynamical hardening}.
\newblock \emph{\mnras}, 474:\penalty0 5672--5683, March 2018.
\newblock \doi{10.1093/mnras/stx3134}.

\bibitem[{Lidov}(1962)]{Lidov1962}
M.~L. {Lidov}.
\newblock {The evolution of orbits of artificial satellites of planets under
  the action of gravitational perturbations of external bodies}.
\newblock \emph{\planss}, 9:\penalty0 719--759, October 1962.
\newblock \doi{10.1016/0032-0633(62)90129-0}.

\bibitem[{Mandel}(2016)]{Mandel2016a}
I.~{Mandel}.
\newblock {Estimates of black hole natal kick velocities from observations of
  low-mass X-ray binaries}.
\newblock \emph{\mnras}, 456:\penalty0 578--581, February 2016.
\newblock \doi{10.1093/mnras/stv2733}.

\bibitem[{Mandel} and {de Mink}(2016)]{Mandel2016}
I.~{Mandel} and S.~E. {de Mink}.
\newblock {Merging binary black holes formed through chemically homogeneous
  evolution in short-period stellar binaries}.
\newblock \emph{\mnras}, 458:\penalty0 2634--2647, May 2016.
\newblock \doi{10.1093/mnras/stw379}.

\bibitem[{Michaely} and {Perets}(2016)]{Michaely2016}
E.~{Michaely} and H.~B. {Perets}.
\newblock {Tidal capture formation of low-mass X-ray binaries from wide
  binaries in the field}.
\newblock \emph{\mnras}, 458:\penalty0 4188--4197, June 2016.
\newblock \doi{10.1093/mnras/stw368}.

\bibitem[{Moe} and {Di Stefano}(2016)]{Moe2016}
M.~{Moe} and R.~{Di Stefano}.
\newblock {Mind your Ps and Qs: the Interrelation between Period (P) and
  Mass-ratio (Q) Distributions of Binary Stars}.
\newblock \emph{ArXiv e-prints}, June 2016.

\bibitem[{Perets} and {Kouwenhoven}(2012)]{Perets2012a}
H.~B. {Perets} and M.~B.~N. {Kouwenhoven}.
\newblock {On the Origin of Planets at Very Wide Orbits from the Recapture of
  Free Floating Planets}.
\newblock \emph{\apj}, 750:\penalty0 83, May 2012.
\newblock \doi{10.1088/0004-637X/750/1/83}.

\bibitem[{Peters}(1964)]{Pet64}
P.~C. {Peters}.
\newblock {Gravitational Radiation and the Motion of Two Point Masses}.
\newblock \emph{Physical Review}, 136:\penalty0 1224--1232, November 1964.
\newblock \doi{10.1103/PhysRev.136.B1224}.

\bibitem[{Petrovich} and {Antonini}(2017)]{Petrovich2017}
C.~{Petrovich} and F.~{Antonini}.
\newblock {Greatly Enhanced Merger Rates of Compact-object Binaries in
  Non-spherical Nuclear Star Clusters}.
\newblock \emph{\apj}, 846:\penalty0 146, September 2017.
\newblock \doi{10.3847/1538-4357/aa8628}.

\bibitem[{Repetto} et~al.(2012){Repetto}, {Davies}, and
  {Sigurdsson}]{Repetto2012}
S.~{Repetto}, M.~B. {Davies}, and S.~{Sigurdsson}.
\newblock {Investigating stellar-mass black hole kicks}.
\newblock \emph{\mnras}, 425:\penalty0 2799--2809, October 2012.
\newblock \doi{10.1111/j.1365-2966.2012.21549.x}.

\bibitem[{Repetto} et~al.(2017){Repetto}, {Igoshev}, and
  {Nelemans}]{Repetto2017}
S.~{Repetto}, A.~P. {Igoshev}, and G.~{Nelemans}.
\newblock {The Galactic distribution of X-ray binaries and its implications for
  compact object formation and natal kicks}.
\newblock \emph{\mnras}, 467:\penalty0 298--310, May 2017.
\newblock \doi{10.1093/mnras/stx027}.

\bibitem[{Rickman}(1976)]{Rickman1976}
H.~{Rickman}.
\newblock {Stellar perturbations of orbits of long-period comets and their
  significance for cometary capture}.
\newblock \emph{Bulletin of the Astronomical Institutes of Czechoslovakia},
  27:\penalty0 92--105, 1976.

\bibitem[{Rodriguez} et~al.(2016){Rodriguez}, {Chatterjee}, and
  {Rasio}]{Rodriguez2016}
C.~L. {Rodriguez}, S.~{Chatterjee}, and F.~A. {Rasio}.
\newblock {Binary black hole mergers from globular clusters: Masses, merger
  rates, and the impact of stellar evolution}.
\newblock \emph{\prd}, 93\penalty0 (8):\penalty0 084029, April 2016.
\newblock \doi{10.1103/PhysRevD.93.084029}.

\bibitem[{Rodriguez} et~al.(2018){Rodriguez}, {Amaro-Seoane}, {Chatterjee},
  {Kremer}, {Rasio}, {Samsing}, {Ye}, and {Zevin}]{Rodriguez2018}
C.~L. {Rodriguez}, P.~{Amaro-Seoane}, S.~{Chatterjee}, K.~{Kremer}, F.~A.
  {Rasio}, J.~{Samsing}, C.~S. {Ye}, and M.~{Zevin}.
\newblock {Post-Newtonian dynamics in dense star clusters: Formation, masses,
  and merger rates of highly-eccentric black hole binaries}.
\newblock \emph{\prd}, 98\penalty0 (12):\penalty0 123005, December 2018.
\newblock \doi{10.1103/PhysRevD.98.123005}.

\bibitem[{Samsing}(2018)]{Samsing2018a}
J.~{Samsing}.
\newblock {Eccentric black hole mergers forming in globular clusters}.
\newblock \emph{\prd}, 97\penalty0 (10):\penalty0 103014, May 2018.
\newblock \doi{10.1103/PhysRevD.97.103014}.

\bibitem[{Samsing} and {D'Orazio}(2018)]{Samsing2018}
J.~{Samsing} and D.~J. {D'Orazio}.
\newblock {Black Hole Mergers From Globular Clusters Observable by LISA I:
  Eccentric Sources Originating From Relativistic N-body Dynamics}.
\newblock \emph{\mnras}, 481:\penalty0 5445--5450, December 2018.
\newblock \doi{10.1093/mnras/sty2334}.

\bibitem[{Sana} et~al.(2014){Sana}, {Le Bouquin}, {Lacour}, {Berger}, {Duvert},
  {Gauchet}, {Norris}, {Olofsson}, {Pickel}, {Zins}, {Absil}, {de Koter},
  {Kratter}, {Schnurr}, and {Zinnecker}]{Sana2014}
H.~{Sana}, J.-B. {Le Bouquin}, S.~{Lacour}, J.-P. {Berger}, G.~{Duvert},
  L.~{Gauchet}, B.~{Norris}, J.~{Olofsson}, D.~{Pickel}, G.~{Zins}, O.~{Absil},
  A.~{de Koter}, K.~{Kratter}, O.~{Schnurr}, and H.~{Zinnecker}.
\newblock {Southern Massive Stars at High Angular Resolution: Observational
  Campaign and Companion Detection}.
\newblock \emph{\apjs}, 215:\penalty0 15, November 2014.
\newblock \doi{10.1088/0067-0049/215/1/15}.

\bibitem[{The LIGO Scientific Collaboration} et~al.(2018){The LIGO Scientific
  Collaboration}, {the Virgo Collaboration}, {Abbott}, {Abbott}, {Abbott},
  {Abraham}, {Acernese}, {Ackley}, {Adams}, {Adhikari}, and
  et~al.]{TheLIGOScientificCollaboration2018}
{The LIGO Scientific Collaboration}, {the Virgo Collaboration}, B.~P. {Abbott},
  R.~{Abbott}, T.~D. {Abbott}, S.~{Abraham}, F.~{Acernese}, K.~{Ackley},
  C.~{Adams}, R.~X. {Adhikari}, and et~al.
\newblock {GWTC-1: A Gravitational-Wave Transient Catalog of Compact Binary
  Mergers Observed by LIGO and Virgo during the First and Second Observing
  Runs}.
\newblock \emph{arXiv e-prints}, November 2018.

\end{thebibliography}

\section*{\label{sec:Appendix-I}Appendix I}

\subsection*{\label{subsec:Post-interaction-orbital-element}Post-interaction
orbital elements}

In order to calculate the post-interaction orbital elements of a pertubed
binary, we apply the following procedure. We first randomize the perturber
trajectory; we randomly sample the closest approach point, $\vec{b}=b\hat{b}$
of the perturber from an isotropic distribution by randomizing the
two spherical coordinate angles, uniformly distributed $\cos\alpha\in\left(-1,1\right)$
and uniformly distributed $\beta\in\left(0,2\pi\right)$, which determine
a plane perpendicular to the closest approach vector $\vec{b}$. The
perturber trajectory can have any direction within this plane, hence
we randomize an additional angle, uniformly distributed in the range
$\xi\in\left(0,2\pi\right)$ as to choose an arbitrary axis in the
plane. Next we randomize $b$ by setting the distribution function
to be $f\left(b\right)\propto b_{{\rm max}}$.

We randomize the state of the binary by randomizing its mean anomaly,
${\scriptscriptstyle M}$ from a uniform distribution between $\left(0,2\pi\right)$
and calculate the post-interaction binary orbital elements, $a_{{\rm post}}$
and $e_{{\rm post}}$ using the impulse approximation. In the impulse
approximation we neglect any motion of the binary during the passage
of perturber. In this approximation we can calculate the velocity
change of each of the components of the binary \citep{Rickman1976,Collins2008}.
The velocity vector change for $m_{1}$ is given by 
\begin{equation}
\vec{\Delta v}_{1}=\int_{-\infty}^{\infty}\frac{Gm_{p}\vec{r}_{p}}{\left|r_{p}\right|^{3}}dt=\frac{2Gm_{p}}{v_{{\rm enc}}b}\hat{b}
\end{equation}
where $\vec{r}_{p}$ is the position of the fly-by perturber set to
be at the closest approach at $t=0$. In the impulse regime we can
approximate the trajectory of the fly-by during the interaction as
a straight line thus
\begin{equation}
\vec{r}_{p}=\vec{b}+\vec{v}_{{\rm enc}}t.\label{eq:approx_trajec}
\end{equation}
The velocity change for $m_{2}$ is then 
\begin{equation}
\vec{\Delta v}=\int_{-\infty}^{\infty}\frac{Gm_{p}\left(\vec{b_{2}}+\vec{v}_{{\rm enc}}t\right)}{\left|\vec{b}_{2}+\vec{v}_{{\rm enc}}t\right|^{3}}dt=\frac{2Gm_{p}}{v_{{\rm enc}}b_{2}}\hat{b}_{2},
\end{equation}
where $b_{2}$ is the closest approach of the fly-by perturber. We
can find the relation between $\vec{b}$ and $\vec{b}_{d}$ in the
straight line trajectory approximation by \citep{Collins2008}: 
\begin{equation}
\vec{b}_{d}=\vec{b}-\vec{r}+\hat{v}_{p}\left(\vec{r}\cdot\hat{v}_{p}\right).
\end{equation}
The change in the relative velocity is simply 
\begin{equation}
\vec{\Delta v}_{r}=\vec{\Delta v_{2}}-\vec{\Delta v}_{1}.\label{eq:DeltaV_relative}
\end{equation}

Given the relative velocity vector change we can calculate the post-interaction
eccentricity vector 
\begin{equation}
\vec{e}\equiv\frac{\left(\vec{v}+\vec{\Delta v}_{r}\right)\times\left(\vec{r}\times\left(\vec{v}+\vec{\Delta v}_{r}\right)\right)}{Gm_{b}}-\frac{\vec{r}}{r},
\end{equation}
to find the post-interaction eccentricity which is the norm of the
eccentricity vector $e_{{\rm post}}=\left|\vec{e}\right|$.

For the post-interaction SMA we calculate the change in orbital energy.
The energy change is due to the velocity kick imparted by the perturber.
In the impulse approximation we model the interaction only via a velocity
change and not by a change in the separation itself during the encounter.
The specific orbital energy is 
\begin{equation}
\varepsilon=\frac{v^{2}}{2}-\frac{Gm_{b}}{r}=\frac{-Gm_{b}}{2a}
\end{equation}
where $r$ is the separation. The velocity change $\vec{\Delta v}_{r}$
can be written as a sum of the parallel (to the instantaneous orbital
velocity) and the perpendicular vectors $\vec{\Delta v}_{r}=\Delta v_{r,\parallel}+\Delta v_{r,\perp}$.
Hence the specific energy change is given by 
\begin{equation}
\Delta\varepsilon=\frac{\left(\vec{v}+\vec{\Delta v}_{r}\right)^{2}}{2}-\frac{v^{2}}{2}=\frac{\Delta v_{r}^{2}+2\vec{v}\cdot\vec{\Delta v}_{r}}{2}\label{eq:delta_E}
\end{equation}
which translates to change in the SMA of 
\begin{equation}
\Delta a=-a\cdot\frac{\Delta\varepsilon}{\varepsilon}
\end{equation}
to give us the final post-interaction SMA $a_{{\rm post}}=a+\Delta a$.

\section*{\label{sec:Appendix-II}Appendix II}

\subsection*{\label{sec:Numrical-Results}Numerical Results}

In order to validate our analytic calculation we numerically verify
equations (21) and (22) as plotted in Figure 1. We calculate the merger
probability for the same binary set up and stellar environment, specifically
for a BBH with $m_{b}=20M_{\odot}$ pertubed by stars of typical mass
$m_{p}=0.6M_{\odot}$ and velocity dispersion of $\left\langle v_{{\rm enc}}\right\rangle =50{\rm kms^{-1}}$
the stellar number density is taken to be $n_{*}=0.1{\rm pc^{-3}}$.
The final merger probability is calculated after $t=10{\rm Gyr}$
since the BBH was initialized. We calculate the GW merger probability
for several SMA values, $a\approx a_{{\rm crit}}=2\times10^{4}{\rm AU}$
and $a=3\times10^{4}{\rm AU},\ 1\times10^{4}{\rm AU}.$ For each value
of $a$ we consider a range of initial eccentricities $e$, and follow
the evolution of each of the systems for $\ensuremath{10{\rm Gyr}}$.
In Table \ref{tab:results} we present the number of systems for each
combination of $\ensuremath{a}$ and $\ensuremath{e}$. For each of
the modeled initial conditions we calculate the fraction of merged
systems in the following way. We record the number of merged systems
$n\left(a,e\right)$ to find the fraction of merged systems out of
the total number of modeled systems 
\begin{equation}
f_{{\rm merged}}=\frac{n\left(a,e\right)}{N\left(a,e\right)}.\label{eq:f_merged}
\end{equation}

We can then also find the standard deviation given by 
\begin{equation}
\sigma\left(a,e\right)=\sqrt{f_{{\rm merged}}\times\left(1-f_{{\rm merged}}\right)\times\frac{1}{N\left(a,e\right)}}.
\end{equation}

\begin{figure}
\includegraphics[width=0.9\columnwidth]{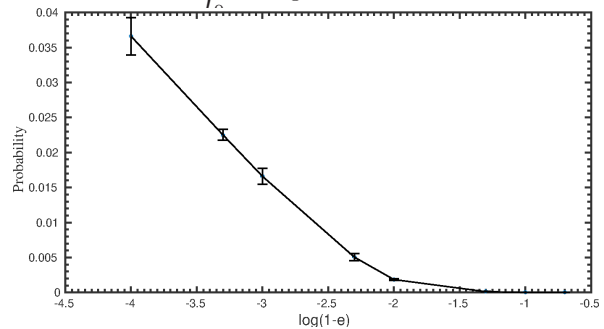}\caption{\label{fig:Probability_a_crit}The estimated merger probability for
BBH with SMA $a=2\times10^{4}{\rm AU}$ as a function of $\log\left(1-e\right)$.
We name this function $f_{{\rm merged}}\left(a,e\right)$ (Eq. (\ref{eq:f_merged})).
In order to calculated the merger probability of the entire ensemble
we weigh each data point with the thermal distribution of eccentricity
(see equation (\ref{eq:Probability_calculation})).}
\end{figure}

In Figure \ref{fig:Probability_a_crit} we present the estimated probability
as a function of initial eccentricity for the case of $a\approx a_{{\rm crit}}=2\times10^{4}{\rm AU}.$

In order to calculate the overall probability, $P$ we weigh $f_{{\rm merged}}$
with a thermal distribution of the initial eccentricities. Analytically
this corresponds to the following integral 
\begin{equation}
P\left(a\right)=\int_{0}^{1}f_{{\rm merged}}\left(a,e\right)\times f\left(e\right)de,\label{eq:Probability_calculation}
\end{equation}
where $f\left(e\right)=2e$ is the thermal distribution of the initial
eccentricities. Numerically we approximate the integral by the following
sum, 
\begin{equation}
P_{1}\left(a\right)=\sum_{i}\frac{1}{2}\left(f\left(a,e_{i+1}\right)+f\left(a,e_{i}\right)\right)\times f_{{\rm merged}}\left(e_{i+1}\right)\times\Delta e_{i},
\end{equation}
where $\Delta e_{i}=e_{i+1}-e_{i}$ and $\max\left\{ e_{i}\right\} <1$.
Another approximation is 
\begin{equation}
P_{2}\left(a\right)=\sum_{i}f\left(a,e_{i}\right)\times f\left(e_{i}\right)\times\Delta e'_{i},
\end{equation}
where $\Delta e_{i}=e_{i+1}-e_{i}$ and $\max\left\{ e_{i}\right\} =1$.
It is clear that $P_{1}$ overestimates the integral while $P_{2}$
underestimates $P$. Hence we take the arithmetical mean of $P_{1}$
and $P_{2}$ and assign it as the calculated probability, $P$. 

In order to calculate the standard deviation we use the following
estimate 
\begin{equation}
\sigma_{{\rm integrated}}=\left(\sum_{i}\sigma\left(a,e_{i}\right)^{2}\times f\left(e_{i+1}\right)\times\Delta e_{i}\right)^{1/2}.
\end{equation}
\begin{table*}
\caption{\label{tab:results}This table summarize the numerical results. In
all runs we chose \textbf{$b_{{\rm max}}=5\times10^{4}{\rm AU}$},o
ensure we simulate the impulsive regime. The stellar number density
is $n_{*}=0.1{\rm pc^{-3}}$ and the velocity dispersion is $\left\langle v_{{\rm enc}}\right\rangle =50{\rm kms^{-1}},$
the perturber mass is $m_{p}=0.6M_{\odot}.$ $n$ - represents the
number of mergers due to GW out of $N$ simulations with the same
initial setup, namely $a$ and $e$. The ratio of $n/N$ is the estimate
of the probability of merger for a specific $a$ and $e$.}

\begin{tabular}{|cc|c|c|c|c|c|c|c|c|}
\hline 
$a\left[10^{4}{\rm AU}\right]$ &  & $e=0.8$ & $e=0.9$ & $e=0.95$ & $e=0.99$ & $e=0.995$ & $e=0.999$ & $e=0.9995$ & $e=0.9999$\tabularnewline
\hline 
\hline 
\multirow{2}{*}{$a=0.5$} & $n$ & - & - & - & $0$ & $34$ & $470$ & $722$ & $1515$\tabularnewline
\cline{2-10} \cline{3-10} \cline{4-10} \cline{5-10} \cline{6-10} \cline{7-10} \cline{8-10} \cline{9-10} \cline{10-10} 
 & $N$ & - & - & - & $24927$ & $11348$ & $24587$ & $24682$ & $24740$\tabularnewline
\hline 
\multirow{2}{*}{$a=1$} & $n$ & - & $0$ & $5$ & $135$ & $472$ & $1801$ & $158$ & $3759$\tabularnewline
\cline{2-10} \cline{3-10} \cline{4-10} \cline{5-10} \cline{6-10} \cline{7-10} \cline{8-10} \cline{9-10} \cline{10-10} 
 & $N$ & - & $100000$ & $250430$ & $99434$ & $100367$ & $94240$ & $7491$ & $125087$\tabularnewline
\hline 
\multirow{2}{*}{$a=2$} & $n$ & $1$ & $3$ & $96$ & $686$ & $126$ & $207$ & $791$ & $183$\tabularnewline
\cline{2-10} \cline{3-10} \cline{4-10} \cline{5-10} \cline{6-10} \cline{7-10} \cline{8-10} \cline{9-10} \cline{10-10} 
 & $N$ & $595746$ & $249568$ & $248101$ & $235553$ & $24495$ & $12474$ & $35122$ & $4999$\tabularnewline
\hline 
\multirow{2}{*}{$a=3$} & $n$ & - & $0$ & $16$ & $204$ & $393$ & $943$ & $327$ & $56$\tabularnewline
\cline{2-10} \cline{3-10} \cline{4-10} \cline{5-10} \cline{6-10} \cline{7-10} \cline{8-10} \cline{9-10} \cline{10-10} 
 & $N$ & - & $99950$ & $99588$ & $96550$ & $94873$ & $94915$ & $24643$ & $2494$\tabularnewline
\hline 
\multirow{2}{*}{$a=4$} & $n$ & - & - & $2$ & $62$ & $85$ & $207$ & $64$ & $61$\tabularnewline
\cline{2-10} \cline{3-10} \cline{4-10} \cline{5-10} \cline{6-10} \cline{7-10} \cline{8-10} \cline{9-10} \cline{10-10} 
 & $N$ & - & - & $50000$ & $48898$ & $24522$ & $29813$ & $7488$ & $4979$\tabularnewline
\hline 
\end{tabular}
\end{table*}

\end{document}